\documentclass[aps,12pt,pra,showpacs,superscriptaddress,preprint]{revtex4}

\usepackage{amsmath}

\usepackage[dvipdfmx]{graphicx}

\usepackage{subfig}

\usepackage{array}

\usepackage{amsfonts}

\usepackage{epsf}

\usepackage{epsfig}

\usepackage{amssymb}

\usepackage{bbm}

\usepackage{delarray}

\usepackage{caption}

\usepackage{amstext}

\usepackage{graphics}

\usepackage{epstopdf}

\catcode`ð=\active

\defð{\u{g}}

\catcode`Ð=\active

\defÐ{\u{G}}

\catcode`Ý=\active

\defÝ{\. I}

\catcode`ö=\active

\defö{\"{o}}

\catcode`Ö=\active

\defÖ{\"O}

\catcode`ü=\active

\defü{\"{u}}

\catcode`Ü=\active

\defÜ{\"{U}}

\catcode`Þ=\active

\defÞ{\c{S}}

\catcode`þ=\active

\defþ{\c{s}}

\catcode`ý=\active

\defý{{\i}}

\catcode`ç=\active

\defç{\d{c}}

\catcode`Ç=\active

\defÇ{\d{C}}

\begin{document}

\title{Thermodynamics Quantities for the Klein-Gordon Equation with a Linear plus Inverse-linear Potential: Biconfluent Heun functions}
\author{\small Altuð Arda}
\altaffiliation[Present adress: ]{Department of Mathematical Science, City University London,
Northampton Square,\\ London EC1V 0HB, UK}\affiliation{Department of
Physics Education, Hacettepe University, 06800, Ankara,Turkey}
\author{\small Cevdet Tezcan}
\affiliation{Faculty of Engineering, Baskent University, Baglica Campus, Ankara,Turkey}
\author{\small Ramazan Sever}
\email[E-mails: ]{arda@hacettepe.edu.tr, ctezcan@baskent.edu.tr, sever@metu.edu.tr}\affiliation{Department of
Physics, Middle East Technical  University, 06531, Ankara,Turkey}

\begin{abstract}
We study some thermodynamics quantities for the Klein-Gordon equation with a linear plus inverse-linear, scalar potential. We obtain the energy eigenvalues with the help of the quantization rule coming from the biconfluent Heun's equation. We use a method based on the Euler-MacLaurin formula to compute the thermal functions analytically by considering only the contribution of positive part of spectrum to the partition function.
\end{abstract}


\maketitle

\newpage

\section{Introduction}

If the one-dimensional linear potential having a form proportional to $|x|$ is considered as the time-like component of a Lorentz vector then this potential becomes related with the Coulomb potential [1, references therein]. The linear potential is also a basic ground for  confinement of the particles having an odd half-integer spin in the view of quantum field theory. If we consider the linear potential as a Lorentz scalar, then it becomes important for the structure of quarkonium. So, the one-dimensional linear potential has received a great interest in literature. The solutions of the Dirac equation and the non-relativistic limit for a linear scalar potential have been studied in Ref. [2]. The bound state solutions of the Dirac equation have been analyzed for the one-dimensional linear potential with Lorentz scalar and vector couplings [1]. Some other relativistic equations such as the Duffin-Kemmer-Petiau [3], and the Klein-Gordon (KG) equation [4] have been also studied for the linear potential. For the non-relativistic case, namely the Schrodinger equation, it is well known that the analytical solutions are obtained in terms of Airy functions [4].

The potential writing as inversely linear ($\sim |x|^{-1}$) denotes another interesting interaction. This is because this potential represents the hydrogen atom in one-dimensional space [5]. The non-relativistic results for this potential show that the ground-state solution has an infinite energy with an eigen function written in terms of delta function near the origin [5]. Analyzing this potential for the Klein-Gordon equation presents unacceptable solutions with the help of continuous dimensionality technique [6]. As a result, it could be interesting to solve the Klein-Gordon equation for the combination of the above potentials writing as
\begin{eqnarray}
V(x)=a_{1}+a_{2}|x|+\,\frac{a_3}{|x|}\,,\nonumber
\end{eqnarray}
to find the statistical quantities for the whole system.

The study of the thermodynamics quantities for quantum systems in different potentials has been received a special interest for last few decades. In Ref. [7], the one-dimensional Dirac-oscillator has been analyzed in a thermal bath, and then the three dimensional case has been computed in Ref. [8]. The Dirac/Klein-Gordon oscillators have been analyzed in thermodynamics point of view by using a different method in Ref. [9]. The Dirac equation on graphene has been solved to study the thermal functions in Ref. [10]. The non-commutative effects on thermodynamics quantities have been also discussed on graphene in literature [11, 12]. The spin-one DKP oscillator has been analyzed for the statistical functions by taking into account the non-commutative effects with an external magnetic field [13]. In Ref. [14], the thermodynamics properties of a harmonic oscillator plus an inverse square potential have been studied within the non-relativistic region.

The paper is organized as follows. In Section II, we obtain the bound state solutions of the Klein-Gordon equation for the above potential with the help of the quantization condition giving the biconfluent Heun's eqution. We will see that the results reveal an energy-eigenvalue equation which is independent of the potential parameter $a_{1}$. In Section III, we compute the partition function, $Z(\beta)$, by using the Euler-MacLaurin formula in terms of a dimensionless parameter $\bar{m}$ by restricting ourselves to the case where the particle-particle interactions appear only. For this case, the partition function does not involve a sum over the negative-energy states [15]. We search then the other thermal quantities such as the free energy, the mean energy, and the specific heat numerically. In Section IV, we give our conclusions.

\section{The Bound States}

The time-independent one-dimensional Klein-Gordon equation with scalar, $V_{S}(x)$, and vector, $V_{V}(x)$, potentials reads as [16]
\begin{eqnarray}
\left\{-\frac{d^2}{dx^2}+Q^2[mc^2+V_{S}(x)]^{2}-Q^2[V_{V}(x)-E]^{2}\right\}\psi(x)=0\,,
\end{eqnarray}
with $Q=1/\hbar c$, $c$ is the speed of light, $m$ is the rest mass, and $E$ is the energy. Here we tend to the vector potential as $V_{V}(x)=0$, and the scalar part given as above. So, we have
\begin{eqnarray}
\frac{d^2\psi(x)}{dx^2}&-&Q^{2}\left[(mc^{2}+a_{1})^{2}+2a_{2}a_{3}+2a_{3}(mc^{2}+a_{1})\,\frac{1}{|x|}+\frac{a^{2}_{3}}{x^2}+2a_{2}(mc^{2}+a_{1})|x|-a^{2}_{2}x^2\right]\psi(x)\nonumber\\&=&-Q^{2}E^{2}\psi(x)\,,
\end{eqnarray}
By defining a new variable $y=\sqrt{Qa_{2}\,}|x|$, and using the abbreviations
\begin{eqnarray}
&&\varepsilon_{1}=\frac{Q}{a_2}[E^2-(mc^{2}+a_{1})^{2}-2a_{2}a_{3}]\,\,\,; A_{1}=-2Qa_{3}(mc^{2}+a_{1})\sqrt{\frac{Q}{a_2}\,}\nonumber\\&&A_{2}=-Q^{2}a^{2}_{3}\,\,\,; A_{3}=-2\sqrt{\frac{Q}{a_{2}}\,}(mc^{2}+a_{1})\,,
\end{eqnarray}
we have
\begin{eqnarray}
\frac{d^2\psi(y)}{dy^2}+\left(\varepsilon_{1}+\frac{A_{1}}{y}+\frac{A_{2}}{y^{2}}+A_{3}y-y^{2}\right)\psi(y)=0\,.
\end{eqnarray}

In order to get a more suitable form for Eq. (4) we write the wave function as
\begin{eqnarray}
\psi(y)=|y|^{p}e^{-qy^{2}-ry}\phi(y)\,,
\end{eqnarray}
with
\begin{eqnarray}
p=\frac{1}{2}+\frac{1}{2}\,\sqrt{1-4A_{2}\,}\,,
\end{eqnarray}
Now substituting Eq. (5) into Eq. (4), the resulting equation reads
\begin{eqnarray}
y\phi''(y)+(2p-2ry-4qy^{2})\phi'(y)+\left[(-4pq+r^{2}+\varepsilon_{1})y-(2pr-a_{1})\right]\phi(y)=0\,,
\end{eqnarray}

This equation is the biconfluent Heun's differential equation having a general form [17]
\begin{eqnarray}
\xi u''(\xi)+(1+c_{1}-c_{2}\xi-2\xi^{2})u'(\xi)+\left\{(c_{3}-c_{1}-2)\xi-\frac{1}{2}\,[c_{4}+c_{2}(1+c_{1})]\right\}u(\xi)=0\,,
\end{eqnarray}
with solutions the so-called biconfluent Heun functions, $HB$
\begin{eqnarray}
\phi(y)\sim HB\left(\sqrt{1-4A_{2}\,},2Q\sqrt{a_{2}\,}(mc^{2}+a_{1}),1+\gamma^{2}+\varepsilon_{1},\frac{4Q^{2}a_{3}}{\sqrt{Qa_{2}\,}}(mc^{2}+a_{1}),y\right)\,.
\end{eqnarray}

The biconfluent Heun's equation has many applications within different subjects to finding the quantization condition and the wave functions for the system under consideration [18-21]. The general solution of this equation can be computed by using the Frobenius methods, and the biconfluent Heun series results in a polynomial form of degree $n$ when [18]
\begin{eqnarray}
\varepsilon_{1}+\frac{1}{4}\,A^{2}_{3}-2p=2n\,,
\end{eqnarray}
with $n=0, 1, 2, \ldots$. By using Eq. (3), we obtain the bound states of the system
\begin{eqnarray}
E^{2}_{n}=2a_{2}a_{3}+\frac{a_{2}}{Q}\left(2n+1+\sqrt{1+4Q^{2}a^{2}_{3}\,}\right)\,,
\end{eqnarray}
with the eigenfunctions
\begin{eqnarray}
\psi(y)&\sim&|y|^{\frac{1}{2}+\frac{1}{2}\,\sqrt{1-4A_{2}\,}}e^{\frac{1}{2}\,(A_{3}y-y^{2})}\nonumber\\ &\times&
HB\left(\sqrt{1-4A_{2}\,},2Q\sqrt{a_{2}\,}(mc^{2}+a_{1}),1+\gamma^{2}+\varepsilon_{1},\frac{4Q^{2}a_{3}}{\sqrt{Qa_{2}\,}}(mc^{2}+a_{1}),y\right)\,.
\end{eqnarray}

We present plots of some eigenfunctions with different quantum number values in Fig. (1). In addition, last two equations makes it possible to handle the single particle level density defined basically as the number of energy levels in the energy interval $dE$ [22], that is,
\begin{eqnarray}
\rho(E)=\frac{dE}{dn}\,,
\end{eqnarray}
which gives for the system under consideration
\begin{eqnarray}
\rho(E)=\frac{Q}{a_{2}}\,\sqrt{E\,}\,.
\end{eqnarray}
where it is clearly seen that the level density depends on the strength of linear and inverse-linear part of potential.

In order to have an equation with same dimensions in the left and right hand sides in (11), let us denote the quantity $"a_{2}a_{3}"$ as $\varepsilon^{2}$ in the rest of computation which makes it possible to write the Eq. (11) more clearly as
\begin{eqnarray}
E_{n}=\mp\, \varepsilon\sqrt{2+q^{-1}(2n+1+\sqrt{1+4q^{2}\,})\,}\,.
\end{eqnarray}
with a dimensionless parameter $q=Qa_{3}$. In the next Section, we compute the thermal functions in terms of a dimensionless parameter $\bar{m}$ written with the help of $\varepsilon$.

\section{The Thermodynamics Quantities}

The partition function given as a summation over all the quantum states can be written as [7]
\begin{eqnarray}
Z(\beta)=\sum_{n=0}^{\infty}e^{-(E_{n}-E_{0})\beta}=e^{\beta E_{0}}\sum_{n=0}^{\infty}e^{-\beta\varepsilon \sqrt{\sigma_{1}n+\sigma_{2}\,}}\,,
\end{eqnarray}
where $\beta=1/k_{B}T$, $k_{B}$ Boltzmann constant, $T$ temperature in Kelvin with the constants $\sigma_{1}=2/q$, and $\sigma_{2}=2+(1/q)(1+\sqrt{1+4q^2\,})$. We tend to compute the following thermal quantities such as the free energy, the mean energy, and the specific heat written in terms of the partition function
\begin{eqnarray}
F(\beta)&=&-\frac{1}{\beta}\,\text{ln}\,Z(\beta)\,,\nonumber\\
U(\beta)&=&-\frac{\partial}{\partial\beta}\,\text{ln}\,Z(\beta)\,,\nonumber\\
C(\beta)&=&-k_{B}\beta^2\frac{\partial}{\partial\beta}U(\beta)\,,
\end{eqnarray}

The following integral equation [7, 8]
\begin{eqnarray}
\int_{0}^{\infty}e^{-\beta_{1}\sqrt{\beta_{2}n+\beta_{3}\,}}dn=\frac{2}{\beta_{1}^{2}\beta_{2}}\,e^{-\beta_{1}\sqrt{\beta_{3}\,}}(1+\beta_{1}\sqrt{\beta_{3}\,})\,,
\end{eqnarray}
shows that the partition function in Eq. (16) is convergent. The result in Eq. (18) makes it possible to compute the partition function with the help of the Euler-MacLaurin formula
\begin{eqnarray}
\sum_{n=0}^{\infty}f(n)=\frac{1}{2}f(0)+\int_{0}^{\infty}f(x)dx-\sum_{i=1}^{\infty}\frac{1}{(2i)!}B_{2i}f^{(2i-1)}(0)\,,
\end{eqnarray}
where $B_{2i}$ are the Benoulli numbers, $B_{2}=1/6$, $B_{4}=-1/30$, $\ldots$ [7, 8]. Up to $i=2$, Eq. (16) with the help of (13) gives the partition function of the system written in a dimensionless parameter $\beta \varepsilon=1/\bar{m}$ as
\begin{eqnarray}
Z(\bar{m})=\frac{1}{2}+\frac{2\bar{m}^{2}}{\sigma_{1}}(1+\frac{\sqrt{\sigma_{2}\,}}{\bar{m}})+\frac{\sigma_{1}}{24\bar{m}\sqrt{\sigma_{2}\,}}-
\frac{\sigma^{3}_{1}}{5760\bar{m}\sigma^{5/2}_{2}}(3+3\frac{\sqrt{\sigma_{2}\,}}{\bar{m}}+\frac{\sigma_{2}}{\bar{m}^{2}})\,.
\end{eqnarray}

We observe that the thermodynamic quantities in Eq. (17) depend the parameter $q$ including the potential parameter. So, we give our all numerical results as the variation of them versus the temperature for three different values of parameter, namely, $q=0.5$, $q=1.0$ and $q=1.5$, in Figs. (2)-(4). Fig. (2) shows that the Helmholtz free energy increase with increasingly value of $a$. In Fig. (3), we see that the effect of the parameter $q$ on the mean energy is more apparent for nearly low temperatures. On the other hand, the plots for different $q$-values for the mean energy are closing to each other. We give the variation of the specific heat according to the temperature in Fig. (4) where it has an upper value while the temperature increases.

Now we give the results briefly for the thermal functions for high temperatures which corresponds to $\beta \ll 1$. For this case, Eq. (20) gives the results
\begin{eqnarray}
&&Z(\bar{m}) \sim \frac{2\bar{m}^{2}}{\sigma_{1}} \sim \bar{m}^{2}Qa_{3}\,,\nonumber\\
&&U(\bar{m}) \sim 2\bar{m}\,,\nonumber\\
&&C(\bar{m}) \sim 2\,.
\end{eqnarray}
where the upper limit for the specific heat can be seen clearly in Fig. (4).

Studying the partition function in Eq. (20) according to the potential parameters shows that the descent contribution, which is inverse-linear, comes from the part of the potential proportional to $|x|$. The other part of the potential proportional to $\frac{1}{|x|}$ gives a weaker contribution, which is linear in some terms and inverse-linear in others. On the other hand, Eq. (21) gives that only the potential parameter $a_{2}$ gives an inverse-linear contribution to partition function while both parameters $a_2$ and $a_3$ give an inverse-squared contribution to the mean energy for high temperature.

\section{Conclusions}

We have obtained the thermodynamics quantities for the Klein-Gordon equation with a linear plus inverse-linear potential by using the quantization condition appeared in biconfluent Heun's equation. The variation of a few eigenfunctions versus spatially coordinate has been given in a figure, and the single particle level density analyzed briefly. The thermodynamics quantities such as the free energy, the mean energy, and the specific heat have been computed by a method based on the Euler-MacLaurin formula. We have obtained the variation of thermal functions according to temperature, and also discussed the results for high temperatures.

\section{Acknowledgments}
One of authors (A.A.) thanks professor A. Fring from City University London and the Department of Mathematics for hospitality. This research was partially supported by the Scientific and Technical Research Council of Turkey and through a fund provided by University of Hacettepe.

The authors also thank the referee for comments which have improved the manuscript.

\newpage

\begin{figure}
\centering \subfloat[][The eigenfunction for $n=0$.]{\includegraphics[width=0.5\textwidth,natwidth=610,natheight=642]{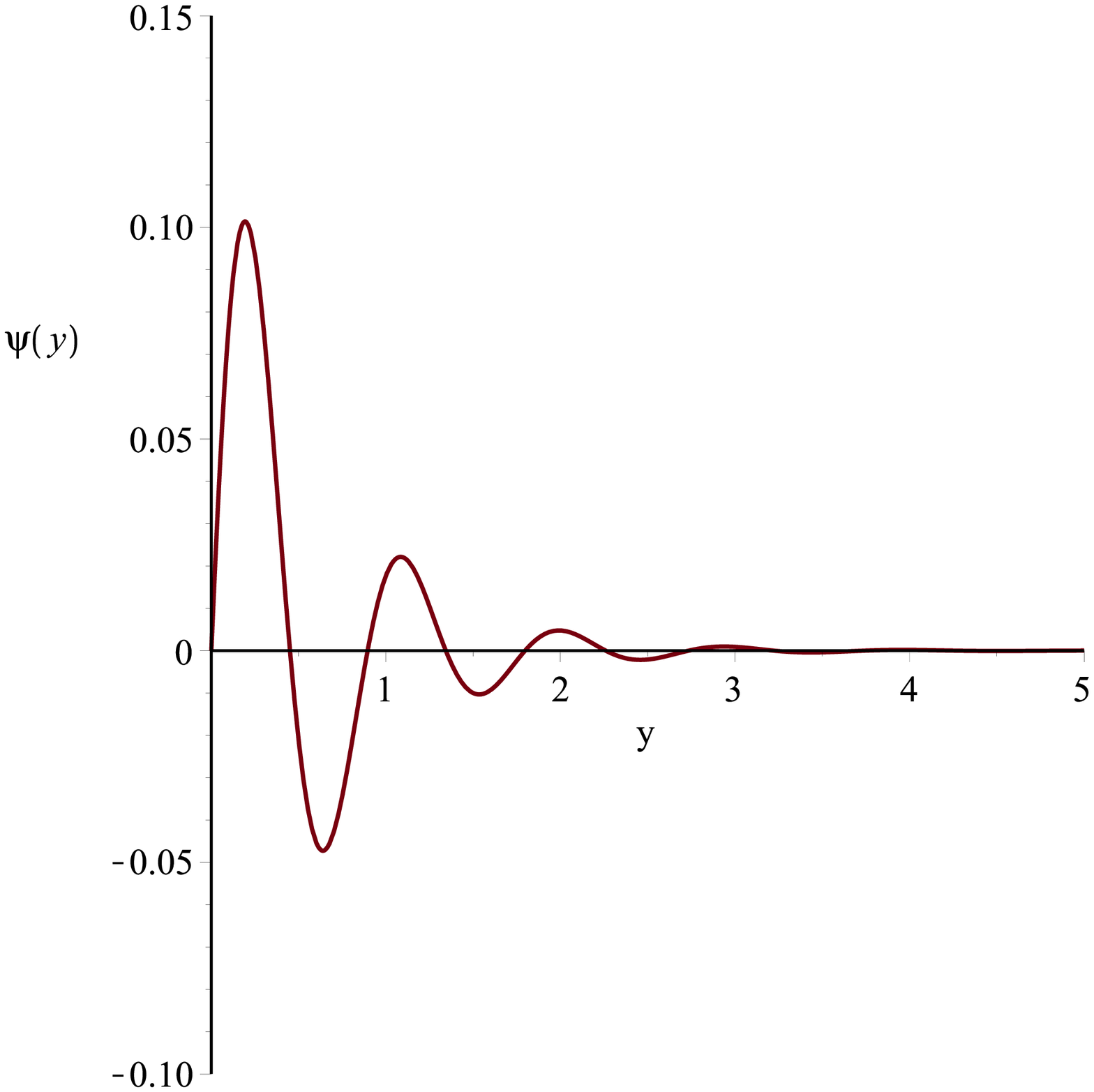}}
\subfloat[][The eigenfunction for $n=5$.]{\includegraphics[width=0.5\textwidth,natwidth=610,natheight=642]{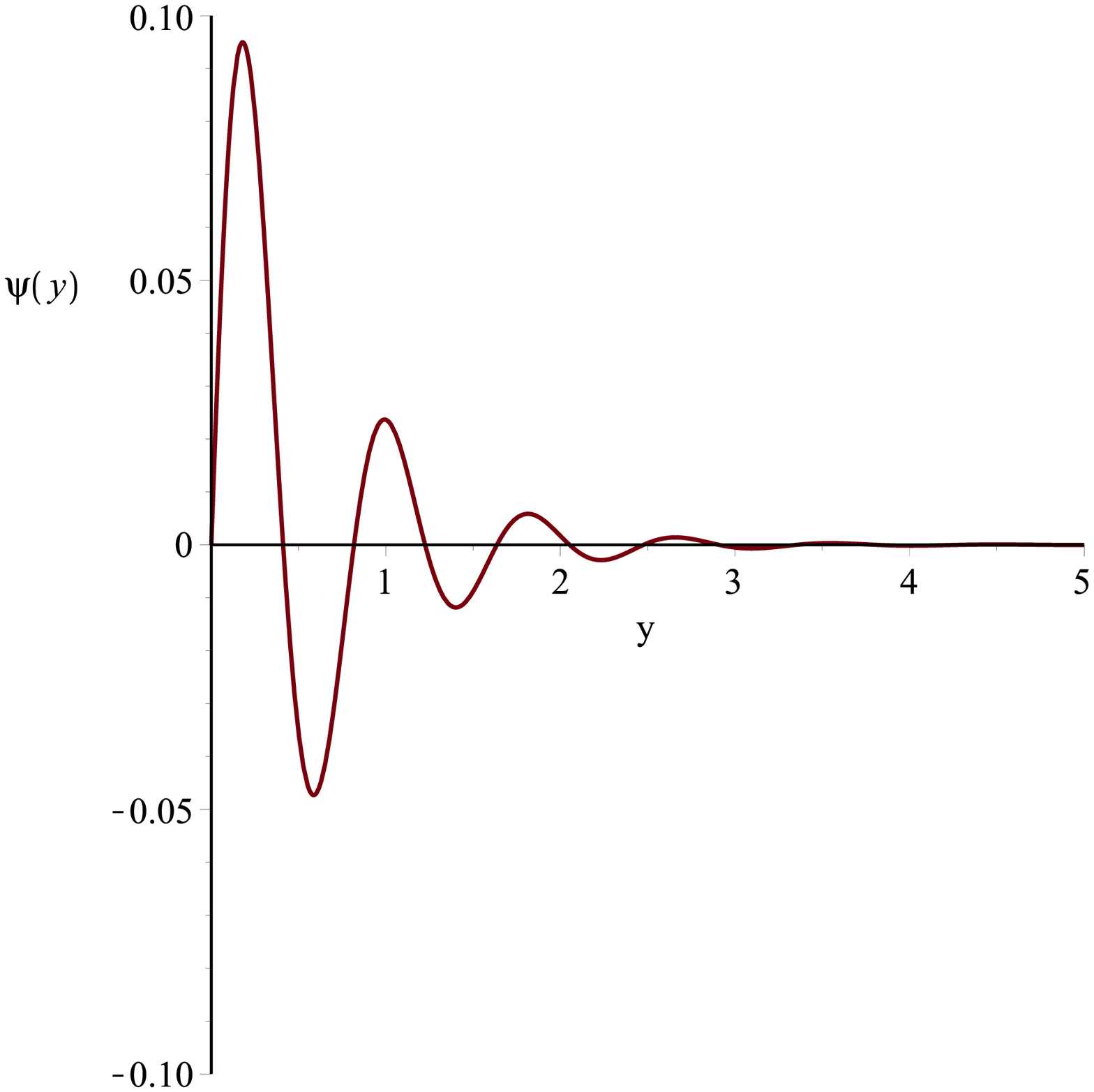}}\\
 \subfloat[][The eigenfunction for $n=10$.]{\includegraphics[width=0.5\textwidth,natwidth=610,natheight=642]{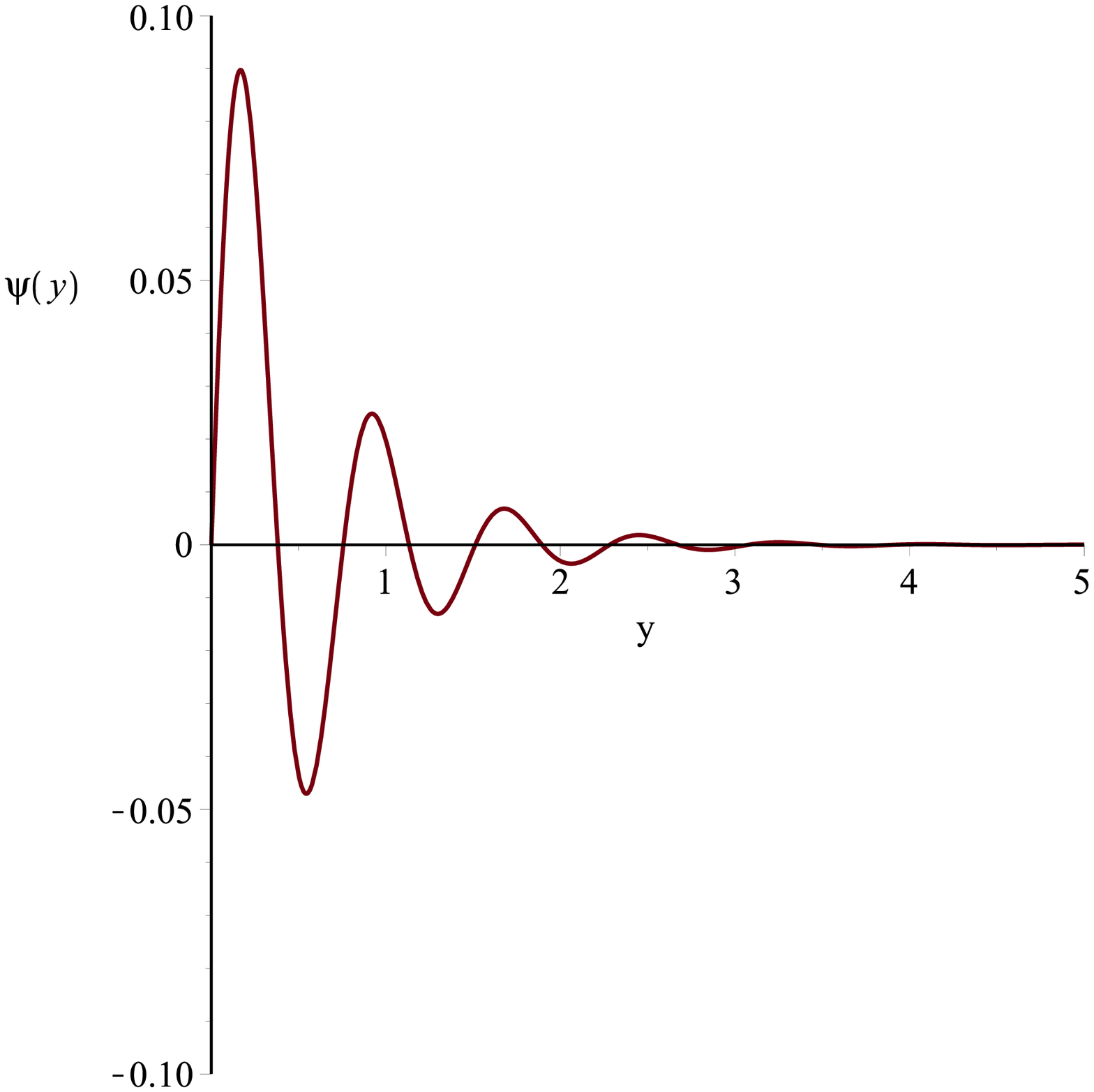}}
 \caption{Some eigenfunctions given in (12) with parameter set $a_1=a_2=a_3=0.1$, $m=0.5$ ($\hbar=c=1$).}
\end{figure}

\newpage

\begin{figure}
\centering
\includegraphics[height=3.5in, width=5in, angle=0]{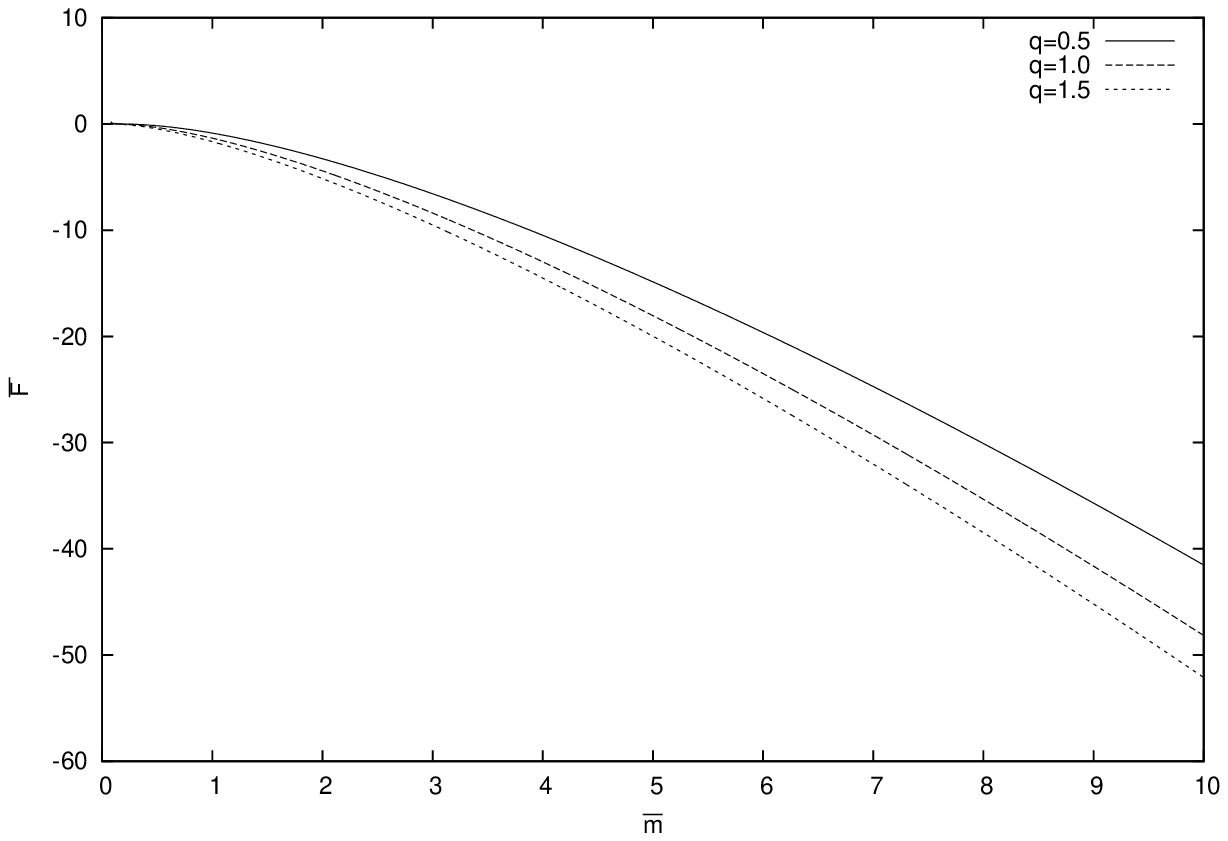}
\caption{The variation of the free energy for the present potential versus $\bar{m}$.}
\end{figure}

\begin{figure}
\centering
\includegraphics[height=3.5in, width=5in, angle=0]{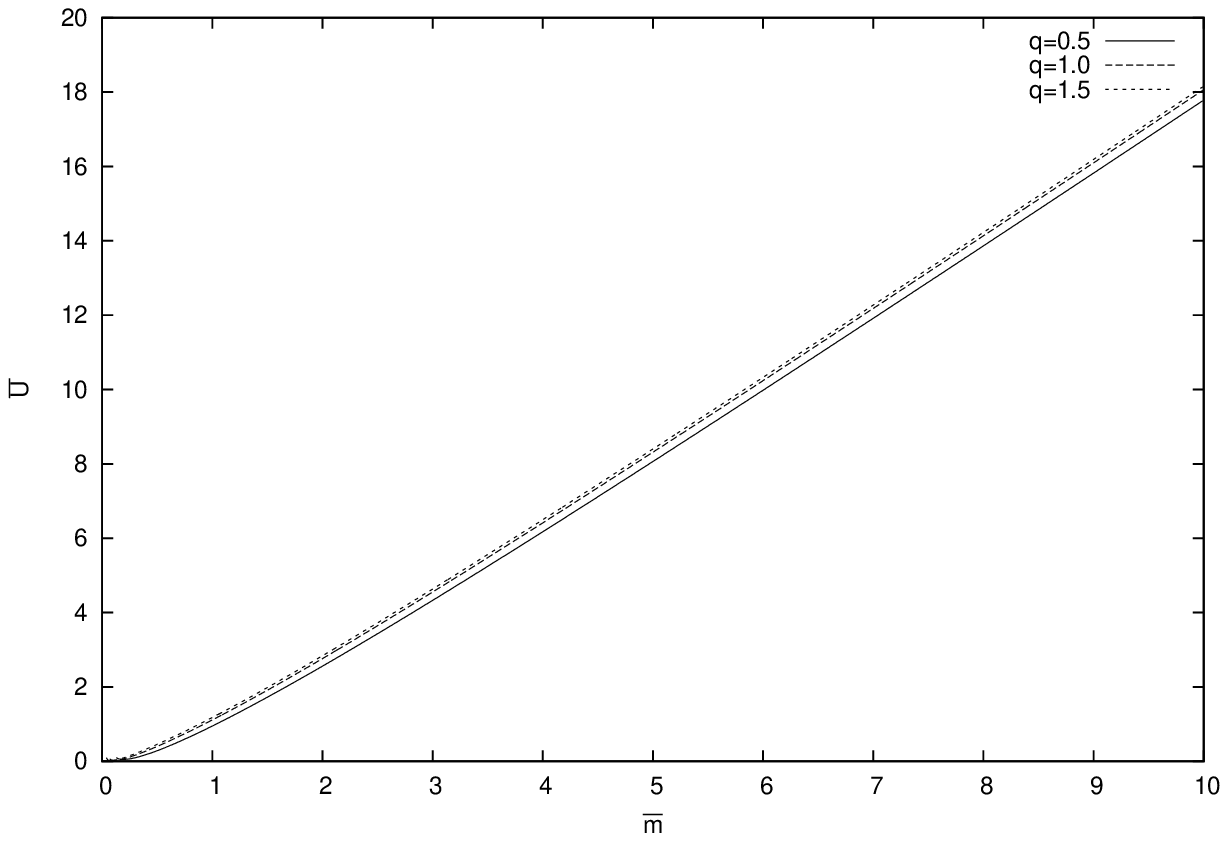}
\caption{The variation of the mean energy for the present potential versus $\bar{m}$.}
\end{figure}

\newpage

\begin{figure}
\centering
\includegraphics[height=3.5in, width=5in, angle=0]{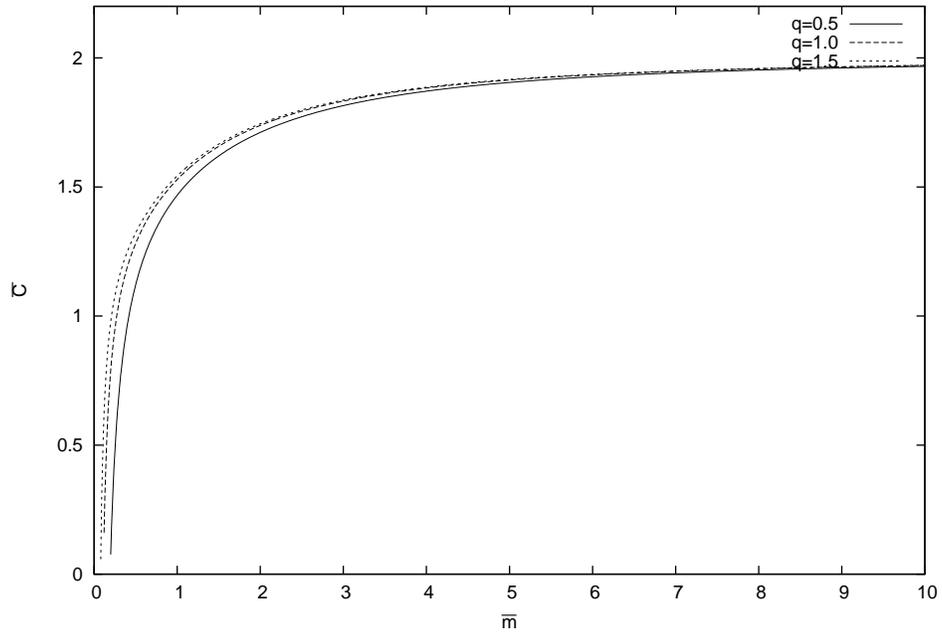}
\caption{The variation of the specific heat for the present potential versus $\bar{m}$.}
\end{figure}

\end{document}